\documentclass[fleqn,twoside]{article} \usepackage{espcrc2}

\usepackage{graphicx}
\usepackage[figuresright]{rotating}

\newcommand{\AmS}{{\protect\the\textfont2
A\kern-.1667em\lower.5ex\hbox{M}\kern-.125emS}}

\title{Status of the PICASSO Project}

\author{M. Barnab\'e-Heider\address[UdM]{Groupe   de    Physique   des
Particules, D\'epartement de Physique, Universit\'e de
Montr\'eal, \\ C.P.6128, Succ. Centre-Ville, Montr\'eal  (Qu\'ebec)  H3C 3J7,
Canada},
E. Behnke\address[IUSB]{Department of Physics and Astronomy, Indiana
University South Bend \\ South Bend, Indiana, 46634, USA },
J. Behnke\addressmark[IUSB], 
M. Di Marco\addressmark[UdM],
P. Doane\addressmark[UdM],
W. Feighery\addressmark[IUSB],
M-H. Genest\addressmark[UdM],  
R. Gornea\addressmark[UdM],
S. Kanagalingam\address[BTI]{Bubbles Technology Industries, Chalk River
(Ontario) K0J 1J0, Canada},
C. Leroy\addressmark[UdM], 
L. Lessard\addressmark[UdM],
I. Levine\addressmark[IUSB], 
J.P. Martin\addressmark[UdM], 
C. Mathusi\addressmark[IUSB],
J. Neurenberg\addressmark[IUSB], 
A.J. Noble\address[QU]{Department of Physics,
Queens University\\ Kingston (Ontario) K7L 3NG, Canada},
R. Noulty\addressmark[BTI],
R. Nymberg\addressmark[IUSB],
S.N. Shore\address[UdP]{Dipartimento   di  Fisica  ``Enrico
Fermi'',Universtit\`a di Pisa, Pisa, I-56127, Italy},
\underline{U. Wichoski}\addressmark[UdM],
V. Zacek\addressmark[UdM]}

\begin{document}

\begin{abstract}
The Picasso  project is a dark  matter search experiment  based on the
superheated  droplet  technique.  Preliminary  runs performed  at  the
Picasso Lab in Montreal have  showed the suitability of this detection
technique  to  the  search  for  weakly interacting  cold  dark  matter
particles. In  July 2002, a  new phase of  the project started. A
batch of six 1-liter detectors with an active mass of approximately 40g
was installed in a gallery of the SNO observatory in Sudbury, Ontario,
Canada at a depth of 6,800  feet (2,070m). We give  a status report on
the new experimental setup,  data analysis, and preliminary limits on
spin-dependent neutralino interaction cross section.
\vspace{1pc}
\end{abstract}

\maketitle

\section{Introduction}

The concept of dark matter has been around for almost a century. 
The most robust indication of the existence of this unique kind of matter, 
which is not susceptible to any interaction other than gravity and
possibly a very weak interaction with ordinary matter, was the
observed inconsistency between the rotation curves of spiral
galaxies and their estimated amount of luminous matter.

The most recent results from WMAP satellite observations of the cosmic
microwave background (CMB) \cite{CMB} together with the observations of
the relation between redshift and distance in type 1a supernovae
\cite{SN} indicate that the total energy density of the Universe is, 
$\Omega_{total} = 1.02 \pm 0.02$ (flat Universe). These results also
indicate that the two main constituents 
of the Universe are the dark energy, a new form of energy that contributes to 
the energy density of the Universe with $\Omega_{\Lambda} = 0.73 \pm 0.02$, 
and matter that contributes with $\Omega_{m} = 0.27 \pm 0.04$. 
However, big-bang-nucleosynthesis predictions (supported by many
observations including CMB measurements) indicate that the density of ordinary
matter (baryon density) in the Universe is, 
$\Omega_b = 0.044 \pm 0.004$. This implies that most of the matter in
the Universe is non-baryonic (dark). 

The most promising non-baryonic candidate from the particle physics
point of view is the neutralino, a weakly interacting massive particle
(WIMP) that appears in many super-symmetric theories. The interaction
of the neutralino with nuclei of ordinary matter is of electro-weak
strength and can be either coherent or spin-dependent. Being a relic
of the super-symmetric phase of the early Universe, the neutralino is
expected to be found in halos of galaxies. In the case of our
Galaxy, neutralinos are expected to have a Maxwellian velocity
distribution with a mean 
velocity of $\sim 300$ km/s. At the Sun's position,
the expected energy density is $\sim 0.3$ GeV/cm$^3$. 

\section{The Experimental Technique}

Picasso detectors \cite{astro} consist of containers filled with
droplets  of a room-temperature superheated liquid (fluorinated halocarbons)
dispersed in a gel. The interaction of a dark matter particle with an
atomic nucleus in the droplets (i.e. fluorine or carbon) is expected
to make the nucleus recoil. If the energy
deposited by this recoiling nucleus within a certain length inside the
droplet is large enough, bubble formation is triggered \cite{seitz} and an
explosive phase transition occurs. The resulting shock waves can be 
detected by piezo-electric sensors. 
 
\section{Picasso at SNO}

In 2002, Picasso started taking data in the water purification gallery
of the SNO underground facility at a depth of 6,800 feet. The setup
consists of six 1-liter detector modules. The total active mass of the
ensemble of detectors is $\sim$ {40 g}. These modules are
divided into 2 groups according to their superheated liquid composition:
3 so-called BD1000 detectors and 3 BD100 detectors. Due to the different
compositions of the superheated liquids, these two groups of detectors
have distinct operating temperature ranges. Therefore, the BD1000 and
BD100 detectors are kept in two separate temperature-control systems
(TCS) so that the temperature of each group of detectors can be
regulated independently. The TCSs are surrounded by water cubes 
for neutron shielding. 

The data acquisition system continuously monitors the piezos' output and
triggers when a given signal exceeds the established minimum amplitude. It
is worth mentioning at this point that, if left undisturbed,  the droplets 
of superheated liquid are stable for indefinite periods of time. 

Due to the nature of the detectors, data acquisition periods must be
followed by a period of recompression so that the continuously growing
gas bubbles, created by the explosive expansion of the superheated
liquid droplets, can be brought back to the original droplet
state. The best results have
been achieved by setting 10 hours of recompression after 30 hours of
data taking. 

The detectors are fine tuned to detect dark matter
particles. However, at the operating temperature range that the detectors
have to be kept to detect dark matter optimally, they are also
sensitive to neutrons and alpha particles, which make up the main particle 
background.  The origin of the background particles can be internal 
or external. 

The external background consists mostly of fast neutrons
coming from the rocks of the walls of the cavern (cosmic ray induced
neutrons are effectively attenuated at the depth of the cavern).
These neutrons are relatively easy to avoid in the mine by surrounding the
detectors with water cubes. The detectors are not sensitive to low-energy 
neutrons (nor to $\gamma$ and $\beta$ radiation) at the
operating temperature range. Also, the setup was  designed to be radon tight.

The internal background consists mainly of alpha
particles produced in the decay chains of uranium and thorium present
as contaminants in the constituents of the detector (and possibly
radon contamination). These alpha particles are very difficult to
remove as the level of contamination of the ingredients of the
detector has to be lower than $10^{-10} - 10^{-11}$g per gram.

Figure 1 shows the data taken with one of
the BD1000 (8g of active mass) detectors in our laboratory in Montreal
and at SNO. The solid line depicts the response of the detector
to the internal alpha background, which is an excellent fit for the data
taken at SNO. The dashed line depicts the sum of the (dominant)
external  neutron background and the internal alpha background. This
clearly shows that the data at SNO is dominated by the internal
background.  
\begin{figure}[htb]
\label{figure:Fig1}
\includegraphics[width=5.0cm,angle=-90]{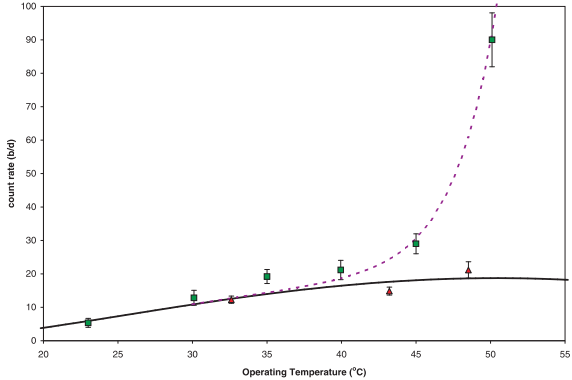}
\caption{Comparison between the data taken in a paraffin shielding 
setup in Montreal (squares) and at SNO (triangles). The 
solid line is the expected alpha response of the detector.}
\end{figure}

Because the detector response is a threshold function \cite{seitz} that is
temperature dependent, data must be taken at different temperatures to
cover the
expected recoil energy spectrum. The data taking is
divided into series that consist of a variable number of 30-hour
sessions and 10 hours of recompression between
them. During a series, the detectors are kept at a constant temperature
that is usually different for the BD1000 and the BD100. 
On an individual event basis, it is impossible to distinguish between the 
signal produced by the expected interaction of dark matter particles with 
the detector and the one produced by background particles. 
Due to that, lowering the overall particle background is
critical.  Besides, the background has to be well
known so that a given enhancement in the count rate (beyond what is
expected to be produced by the background) might be an indication of
the interaction of dark matter particles with the detector. 

Our data is consistent with an internal alpha particle background as
shown by the measurements and Monte Carlo simulations \cite{como2003}.  
Figure 2 shows the response of one of the BD1000 detectors to
background measurements. Until November 2003, $\sim$ 1,700 effective
hours of data were taken at SNO. Due to the large spin-dependent cross
section of the fluorine (present in the superheated  liquid), we
are able to obtain competitive results (for the amount of active mass). 
For a $M_{\chi} = 50$ GeV/c$^2$ neutralino, cross section, $\sigma =
5pb$ is excluded at $90\%$ C.L.. 

\begin{figure}[htb]
\label{figure:Fig2}
\includegraphics[width=7.0cm,angle=0]{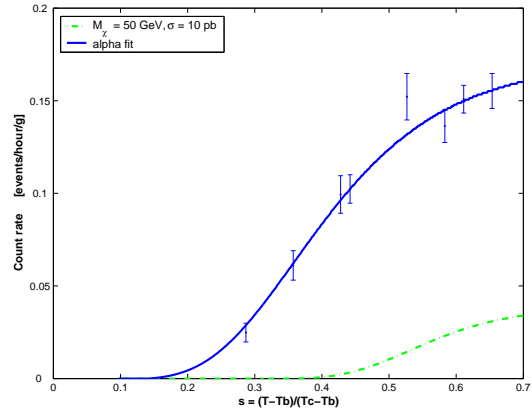}
\caption{Response of a 7g-active-mass BD1000 detector at SNO as a
  function of the temperature expressed in terms of the 
  reduced superheat, \textit{s} \cite{como2003}. The dash-dot line
  depicts the expected response of the detector to 10 pb, 50 GeV
  neutralinos. The solid line is a fit of the expected alpha response
  of the detector to the data.}
\end{figure}

\section{Conclusions}

The setup in operation at SNO has proven to produce data
of excellent quality. The main limitation is the internal alpha
background due to radioactive contaminants in the ingredients of the gel. 
Presently, lowering the internal background is one of the main
efforts of the collaboration. The plans for the future include the 
fabrication of larger detectors and the increasing of the active mass
(a new capillary technique is in R\&D stage). As an intermediate step to
the fabrication of large detectors, a 10-liter prototype is expected to be
ready in the beginning of 2004.

\end{document}